# Grid Computing: The Next Decade

*Report and Summary*

**Program Committee Chair:**
  Jarek Nabrzyski    University of Notre Dame, USA
**Local Chair:**
  Krzysztof Kurowski    Poznan Supercomputing and Networking Center, Poland
**Co-Chairs**:
  Daniel S. Katz    University of Chicago & Argonne National Laboratory, USA
  Andre Merzky    Louisiana State University, USA
**Representatives from Funding Agencies:**
  Gabrielle Allen    NSF, USA
  Konstantinos Glinos    DG INFSO, European Commission
  Ed Seidel    NSF, USA

## Abstract

The evolution of the global scientific cyberinfrastructure (CI) has, over the last 10+ years, led to a large diversity of CI instances. While specialized, competing and alternative CI building blocks are inherent to a healthy ecosystem, it also becomes apparent that the increasing degree of fragmentation is hindering interoperation, and thus limiting collaboration, which is essential for modern science communities often spanning international groups and multiple disciplines (but even 'small sciences', with smaller and localized communities, are often embedded into the larger scientific ecosystem, and are increasingly dependent on the availability of CI.)

There are different reasons why fragmentation occurs, on technical and social level. But also, it is apparent that the current funding model for creating CI components largely fails to aid the transition from research to production, by mixing CS research and IT engineering challenges into the same funding strategies.

The 10[th] anniversary of the EU funded project 'Grid Lab' (which was an early and ambitious attempt on providing a consolidated and science oriented cyberinfrastructure software stack to a specific science community) was taken as an opportunity to invite international leaders and early stage researchers in grid computing and e-Science from Europe, America and Asia, and, together with representatives of the EU and US funding agencies, to discuss the fundamental aspects of CI evolution, and to contemplate the options for a more coherent, more coordinated approach to the global evolution of CI.

This open document represents the results of that workshop — including a draft of a mission statement and a proposal for a blueprint process — to inform the wider community as well as to encourage external experts to provide their feedback and comments.

# 1 - Introduction

Grid computing has been promoted for more than 10 years as the Advanced Computational Infrastructure of the future, see Figure 1. Many scientists and others have considered grid computing as one of main sources of the impact that scientific and technological changes have made on the economy and society. This claim is based on the observation that the usage of large data volumes has become increasingly important to many disciplines, from natural sciences, engineering to the humanities and social sciences. However, despite significant investments in the grid concept, the number of users is not increasing. Instead, new concepts (or at least new terms) like cloud computing seem to be replacing the grid computing approach (or name).

Many fields have depended on computational science simulations, and many now are beginning to depend on computationally intensive data analysis. Infrastructure providers seek to build computational systems that support these researchers. Developing the common vision that is needed to support these efforts is the eventual goal of the Zakopane workshop (Grid Computing: The Next Decade – http://www.gridlab.org/Meetings/Zakopane2012/), which brought together international leaders and early stage researchers in grid computing and e-Science from Europe, the USA, and Asia to discuss the needs and processes towards this goal.

The program offered keynote talks by grid and cloud computing prominent thinkers as well as presentations of modern distributed computing infrastructures, capabilities and applications from users and community leaders. The meeting also served to build research and collaboration bridges between European, US, Asia-Pacific and other region's research organizations. Moreover, representatives from US and European funding agencies discussed funding opportunities for cross-boundary, global collaborations.

Whatever is said about grid computing, it is still a key element of global cyberinfrastructure. The largest scientific computational collaborations, such as Large Hadron Collider collaborations (CMS, ATLAS, ALICE, LHCb), LIGO and GEO600, etc. have deployed and depend on grid computing infrastructures as their production computing engines. Workshop organizers drove discussions around the process by which a national, continental, or global grid computing vision might be established, including discussing the following questions:

- What are the global scientific problems we need to work together on (climate change, EarthCube cyberinfrastructure, emissions, new energy sources, …)?
- Who are the key stakeholders who need to be involved?
- How the global community should move towards truly integrating grid elements at all levels, including those of individual investigators, campuses, countries, and regions?
- How do we move towards integrating networks, clusters, supercomputers, grids, and clouds?
- How do we provide integrated support and training for users?

- How do we integrate software and middleware across this variety of systems?
- How do we provide sufficiently simple abstractions that developers can write applications once, and run them anywhere?
- How do we unify authentication and accounting?
- And underlying all of these decisions, what metrics do they seek to maximize in these processes, and how will they measure them?

Answering these questions is critically important to many science domains, and while the workshop was not able to answer them completely, we believe that this workshop and this subsequent report make a significant contribution toward establishing a common vision among the community leaders who will have to do the work to unify the field, and among the government agencies who will be asked to fund the activities.

Before the workshop a set of relevant documents and white papers have been collected to help participants to answer above-mentioned questions. The EU GridLab project website itself contains many interesting and still fresh ideas for the software development of application tools, middleware services and testbeds for Grid environments driven by end-users [1]. Surprisingly, after ten years, identified research areas in Grid Computing within GridLab reasonable well address challenges that we have to be addresses today, including high-level application APIs, resource management, security, big data, data analysis and visualization, monitoring and information services, science gateways and mobility of end users. It also shows how complex research topics are and how long it takes to produce mature and reliable software to solve problems from mentioned areas in Grid Computing.

For instance the Cyberinfrastructure Framework for 21st Century Science and Engineering (CIF21) presents a set of key complementary and overleaping components that are critical to effectively address and solve the many complex problems facing science and society [2,3]. Core components include data, software, campus bridging and cybersecurity, learning and workforce development, grand challenge communities, computational and data-enabled science and engineering, and scientific instruments. However, only the grand challenge communities have been highlighted in the overall picture, whereas in our opinion representatives from Big Science and Long Tail Science communities have to take an active role in the process of defining requirements for advanced computing infrastructures in the next decade as it is presented in Figure 1.

Researchers representing Big Science or Long Tail Science communities involved in European Strategy Forum on Research Infrastructures (ESFRI) projects have generated many interesting requirements. Key resources, facilities and capabilities that are required from future CIs are listed in the report [4].

In the recent e-IRG White Paper a user-centric approach in the future e-Infrastructure ecosystem has been emphasized [5]. However, the leading edge users from Grand Challenges Communities were mostly identified as key players in setting the strategy for CIs.

In the Strategy for the UK Research Computing Ecosystem document authors have identified specific requirements of scientific communities, ranging from the research-group level, to university, regional and national facilities [6]. Importantly, one of key recommendations from funding agencies was the urgent need for investments in research computing as the current approach is hardware centric, short-term and it does not recognize the critical and long-term role that software and people will play in the future CIs.

Software as a core component of future CIs is also highlighted in Figure 1 and was addressed during the workshop.

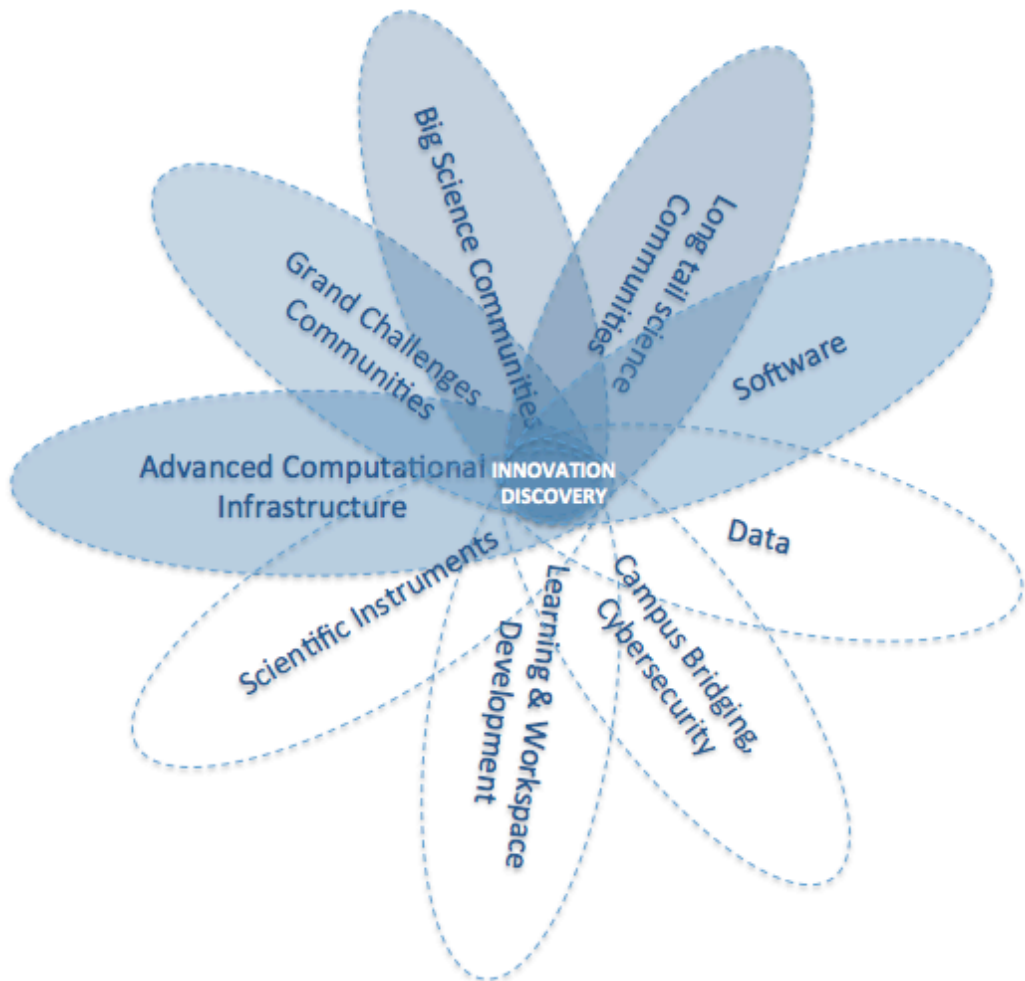

*Figure 1. NSF Cyberinfrastructure Framework for the 21st Century (CIF21), extended by two components representing Big Science and Long Tail Science Communities.*

## 1a - Process and participants

The workshop had a mix of plenary sessions and breakout groups. Plenary sessions included thought-provoking keynote presentations by Peter Coveney, Ian Foster, Miron Livny, Edward Seidel, and Hai Zhuge, invited talks by Ewa Deelman, Rion Dooley, Ian Fisk, Cees de Laat, Maciej Malawski, Karolina Sarnowska-Upton, Shantenu Jha, and Douglas Thain, position statements from Daniel S. Katz, Cees de Laat, Steven Newhouse, Ariel Oleksiak, Morris Riedel, and Satoshi Sekiguchi, and general discussions, including wrap-ups of breakout sessions and discussions of the overall progress of the meeting. The breakouts were focused on three topics: large science, small science (long-tail), and grand challenge communities. Each breakout group met three times, first to address how to gather requirements (§2), second how to develop a process to move forward (§3), and third to discuss actions that could be started immediately (§5).

The Big Science breakouts had the following participants: Rion Dooley, Ian Fisk, David Hart, Tomasz Kuczynski, Miron Livny, Andre Merzky, Radu Prodan, Juliusz Pukacki, Morris Reidel, Miroslav Ruda, Satoshi Sekiguchi, Nour Shublaq, and John Towns (first and second sessions only).

The Small Science (Long-tail) breakouts had the following participants: Piotr Bała, Ewa Deelman, Lukasz Dutka, Ian Foster, Geoffrey Fox, Daniel S. Katz, Jacek Kitowski, Robert Klinc, Bartosz Lewandowski, Daniele Lezzi, Steven Newhouse, Michela Taufer, John Towns (third session only), Von Welch, Ramin Yahyapour, and Hai Zhuge.

The Grand Challenge Communities breakouts had the following participants: Peter Coveney, Cees de Laat, Ewa Deelman, Kostas Glinos, Shantenu Jha, David Jones, Paul Lukowicz, Ariel Oleksiak, Judy Qiu, Karolina Sarnowska-Upton, Edward Seidel, Jarek Nabrzyski (remotely) and Douglas Thain.

# 2 - Status of CIs and requirements

The three breakout groups were asked to address:
1. What are the science requirements for particular community? (big science, long tail science, grand challenge communities)
2. How can we define a "blueprint" or "architecture" to provide an enabling CI for global science in the next decade?

They also tried to define their own identity, in order to get on a common footing and to start the discussion.

## 2a - Big Science

What makes Big Science Big? The large project budget? Big size of the collaboration in terms of the number of people involved? The extremely large amount of data generated? The large complexity of instruments and the large complexity of the required oversight? The long timescales? The answer is 'yes' to all these questions. A common property though is, if compared to Small Sciences / Long Tail, that there is generally a more structured approach to work with end user requirements (simply because nothing else scales). Also, it implies at least some long term planning.

It seems, however, that long-term-planning is often just wishful thinking, as there is no long-term commitment on CI or funding level (even if some define big science as long-term). In general though, there are attempts to collect requirements before Big Science projects start, e.g. with a series of workshops etc., and to stay tightly involved with the user community while evolving the CI stack.

An exception to the above are Big Science communities who run their own CI stack, as they are then often provided with long term funding guarantees. Those setups are not typical. Also, the respective communities seem to re-invent many parts of the CI stack, to ensure its fitness for the target use cases; there is a (perceived or real) assumption that existing components are not able to deliver that.

**Long-term Planning vs. Long-term Strategy:** Planning is what funding agencies ask of scientists, strategy is what scientists ask of funding agencies.

## 2b - Small Science / Long Tail

**What is Small (long-tail) science?**

"Small" or "long tail" science communities were defined as having elements of the following: small collaborations, ad hoc or spontaneous; parts of a large community (e.g., Biology); multidisciplinary; small project size (e.g., "98% of NSF grants are $1m or less"); minor planned role of information technology; democratized access (citizen science).

Examples of long-tail science are in sequencing, where a scientist buys a next-generation sequencer in a lab, does sequences with some postdocs and lab personnel, and is on their own in managing the data; earthquake simulation; pharma simulations; integration advance microdynamic algorithms; etc.

**How to determine the CI needs for long-tail science?**

This could be done by going through a bunch of examples, pulling out requirements, and then looking to see which are common. It could also be done by looking at what scientists are doing and how they are doing it, then looking at what they say they want to do and what they are missing that would enable them to do these things. For the sequencing user, for example, these include: data movement; metadata generation; data storage; data search; provenance; sharing; tagging; and analysis.

Similar studies have been done previously, and rather than duplicating them, we should look at their outputs. For example, two completing teams did this for the NSF XD solicitation, and then merged their gathered requirements into one set [7].

**What are the challenges for long-tail science?**

Amongst the challenges faced by the long-tail scientists is software integration/composability, meaning that the scientists have existing packages they are used to and that they want to use with other software, or in a larger context. They also need to customize software/tools they find for their problems, as well as needing to install such software/tools on the infrastructures on which they are going to run their applications.

The fact that the long tail is so diverse poses the question if one can provide services that are useful to "1.8m" users? (This 1,800,000 is from a slide presented by Steven Newhouse, which stated that there are 1,800,000 researchers funded by the EC, and of these, only about 10,000 are using production cyberinfrastructure such as EGI or PRACE.) How does one know what those services that would be generally useful are? What services would it make sense to provide them centrally? Do we know how to scale services to that level? How many different services would we provide, and which of them would be useful to which users?

This is complicated because we believe that at some level, users do not know what they want, at least no in terms of services. They know what they want to accomplish, but not necessarily what services are needed to do this. Additionally, many things change quickly these days, such as the computing paradigm, desktop tools, science domains and problems, etc., and we need support and react to those changes. Working with the long-tail science users can be a benefit here, as small science is more agile than large science; it can adopt new technologies more quickly and in a less structured manner.

**What are missing capabilities?**

There are two separate groups of things: first, doing the same things they are doing now but at 100 to 100,000 times the rate, and second, doing things that cannot be done today. To some extent, one needs to educate the long-tail science community, to make them computationally sophisticated so they can better understand and explain what they want to do.

Web access is popular because it enables scaling via ease of use. This general idea, to emphasize the interface rather than capability behind the interface, is important. For many scientists, simplicity of the interface is generally important, whether through a web client or other interfaces. This led to the idea of providing simple building blocks as another general principle.

If we look closer at what is missing in local, national and global cyberinfrastructure, we will quickly notice that tools supporting distributed collaborations, data semantic integration, data lifecycle management tools and governance models and policies are still missing.

## 2c - Grand Challenge Communities

Grand Challenge Communities were defined as the next generation of "Grand Challenge Projects" addressing global scientific problems that are too large and diverse even for a consortium of research groups. Example problems include:

- Modeling and understanding of gamma ray bursts requiring many different disciplines (gravitational physics, astrophysics, chemistry, mathematics, computer science)
- Particle physics, gravitational science, or astronomy motivated by large instruments such as the Large Hadron Collider, LIGO/VIRGO, LSST, etc.
- Understanding the human brain

The focus of the sessions at this workshop understood what is needed to support these communities, not necessarily specific to the grand challenges themselves, but the incremental needs to support the research communities. One particular focus was around improving the access to existing research, data and instruments, for example improving access for students. Four aspects were identified related to access:

1. Technical infrastructure
2. Policies such as intellectual property management and OpenAccess
3. Social and cultural interactions and understanding how communities cooperate
4. Regulations for data, such as data privacy

Universities and institutions were seen to have a responsibility to play a role, for example at PSNC the data services team assist affiliated universities with services such as optical networks, identity management, videoconference/HDTV and broadcast capabilities, cloud capabilities including license centralization, data archive and back up.

For research projects, policies are needed that encourage data to be made available and define validation standards, and then there needs to be places to put the data where it can be linked to publications, as well as powerful search engines to find the most relevant and useful data. Systems supporting the activity of managing the use of data from its point of creation to ensure it is available for discovery and re-use in the future are needed. But research data lifecycle management systems are hard to be implemented, especially for grand challenge communities that have a long history of operations, and never addressed this issue. HEP community for example has just started some planning activities oriented towards developing data preservation solutions.

The summary of points from the grand challenge communities sessions were:

1. Supporting the progressive advancement of scientific work
2. Management and incentivation of social interactions
3. Technical data storage and identification
4. Access to consistent computing resources and core data services
5. Funding model and governance for global research communities
6. Training of students

## 3 - Process

This section discusses how the three different breakout groups tried to address the following questions:

1. How would we develop a (minimal) high level blueprint/framework/conceptual architecture or set of processes (or is there a better word?) to organize and coordinate the development and support of cyberinfrastructure, e.g. could expect that this would include

- minimal security assurances, identity management
- data sharing policies
- collaborative software development
- campus bridging to international infrastructures
- governance mechanisms

- continued innovation, as illustrated by the rapid progress of commercial offerings
- reuse and best practices

2. How would these processes aid in activities such as

- sustainability
- international cooperation
- any others?

3. How to turn this "blueprint" into a set of actionable processes?

## 3a - Big Science

**How to develop the "blueprint" process?**

There is strong (but not universal) agreement that adoption of existing solutions is preferable to implementing new solutions - but that does not go well in the current funding structure (funding agencies are funding solutions, not processes [Earthcube is an exception here.]). The large international CIs may be able to coordinate on that level (EGI, PRACE, NAREGI, XSEDE), but user communities will likely only participate when there is an obvious benefit, not before. Note that culture differences are also hindering a more global coordination (e.g. humanities).

Ideally, the blueprint process will support adoption of existing solutions, and thus greatly increase coherence and sustainability of the CI stack. While it is hard to distinguish needs from desires, the blueprint needs to focus on needs in order to be manageable and efficient. An open process for a global CI blueprint can obviously get unwieldy, and multiple funding agencies may complicate things further - but a clear demonstration on return-of-investment (which is difficult though) and a clear process will help to consolidate. A blueprint would, without a doubt, pose a very significant amount of work, but there is agreement that this effort is worthwhile, and also economic. Nevertheless, it requires significant commitment from the key stakeholders.

There is consensus that a global blueprint process is needed – even if 'global' is not acceptable for some communities. While funding agencies are amongst the stakeholders, they should not own the blueprint process, but support it, implicitly and explicitly. The process would be most efficient if owned by a relatively small group with global input -- which is a difficult balance.

## 3b - Small Science / Long Tail

**How to develop the "blueprint"?**

We could either use standards (defining common interfaces) or use policies. The standards approach has been tried in OGF [8], and while this work has been useful, it did not completely succeed. There seem to be two reasons why. First, in order for a standards approach to be successful, there needs to be multiple people who are active in projects where the standard would be applied, and these people need to want to define a standard. At OGF, there have been times where people have participated in OGF because their project said they would, but it wasn't clear that they wanted the effort to be successful. Second, a standard cannot be developed in isolation, and operational issues related to use of the standard need be to part of the discussion. In the case of OGF, there was at times a separation between people who wanted to develop a standard and people who would then need to implement it in a working system.

**How to determine what services are useful and what should be supported?**

This was an area where one could make a distinction between big science and long-tail science, as big-science seemed to create infrastructures in a top-down manner, but long-tail science did things more bottom up. Specifically, if one could create an ecosystem of software and services, one could then use a "market"-based approach, where multiple tools and services can be developed, and one could see which people use and which gain traction. This is similar to the commercial and open-source community today.

For example, some project in the earthquake community spent 1 1/2 years discussing needs and then 6 months developing tools for that community, with a successful outcome. The overall result was that these two years of work made the users' analyses run faster.

**Defining a high-level architecture, as well as defining the technical process for doing so.**

Based roughly on the EGI model presented by Steven Newhouse, we decided to plan a process that is cyclic at two (or possibly more) levels.

In this process, the initial step is to define a base set of interfaces that will be exposed from the underlying hardware systems. This can be thought of as defining a hardware platform. In parallel, a set of common services needs to be defined that all instances of the hardware platform can use. (For example, identity management is a likely common service.) Instances of the hardware platform could compete with each other, as could instances of the services, though at some point, ideally the services would settle on the best instances, which would continue forward.

Next, groups can define platforms on top of this hardware platform, and possibly using these common services, with the goal of supporting communities. These platforms can be seen as smaller instances of the previous hardware platform, with each having a set of common services. For example, there might be an astronomy platform with a set of astronomy services. Or there could be a Hadoop platform, with possibly a set of common Hadoop services.

Next, the two levels of cycling continue. The infrastructure platform can be improved, and a new level on top of the new community platforms can be built, again possibly with common services. There might be a platform designed to support data-intensive astronomy, for example. At each level, specific applications could also be developed, so an astronomy application could sit on the hardware platform and the astronomy community platform, or it could sit directly on the hardware platform.

No matter which specific platforms are developed, it's important to make sure that the platforms are always able to track advances in the underlying technologies, which often could be changes of the platforms under the agreed upon interfaces, but sometimes will require changes to the interfaces as well. The key is an understanding that nothing is completely static, but most interfaces shouldn't change very often.

It is also important to clearly define what is common between infrastructures, and what is not, both for the common services, and also for the common interfaces above the platforms.

There are certainly a number of possible issues with this model. One is governance. As learned from OGF, agreements on architecture and interfaces ideally are binding on the participants in the processes. In this case, it's not clear how to make this happen. Perhaps funding agencies need to be involved in the process, and they could require that the output is binding on the groups they fund? Or perhaps the initial versions are optional, with an expectation that later cycles would become binding? More work and discussion is needed to resolve this.

Another issue is related to the competition between implementations of platforms and services. The group liked the idea of an "app store" that would track usage of implementations, offer reviews and success stories, etc., at least for components that are sufficiently low-cost or high-risk. However, other components (e.g., those that are high-cost and/or low-risk) might be better with coordinated or shared development of a single version. For this latter group of components, stakeholders need confidence that the shared/coordinated software will really appear, do what they want, and will continue to exist.

The "app store" concept leads to questions about the number of app stores: would there be one or many? The group thought that more than one might be needed today for various architectures (much like smartphone apps have iOS and Android versions), but this is desirable in the long term. Ideally, sufficient standards or a common interface to infrastructure would allow there to just be one app store, at least in appearance to the users. There could actually be multiple app stores, perhaps federated, that are presented as a single store, perhaps hiding apps that won't work on a particular infrastructure or platform. With multiple app stores, however, it would be important to have a single trusted point for usage metrics for those apps.

**How to turn blueprint into actionable processes?**

A partial answer from EGI is to start with an architecture workshop, where infrastructure providers would discuss a common interface they could provide/define on top of their infrastructures, followed by a software workshop (for example, based on the ScienceSoft project: http://sciencesoft.web.cern.ch) where software provides could define interfaces they would offer on top of their software, which would sit on top of the provider interfaces. These groups could iterate, with more (higher) levels coming in during later iterations.

It is clear that neutral governance is needed, where all groups can have a say without any one dominating and making decisions. We did not come to any decisions on who would be invited to these workshops. Nor did we reach a conclusion of where users/scientists would fit in this process.

**Summary:**

1. Focus on solutions that are relatively simple individually, and which can be built into more by combining multiple of these solutions. On top of such general building blocks, one could imagine additional layers that are less general, perhaps customized for a community, such as a science domain or a technology like Hadoop. Steven Newhouse presented a slide that discussed "Community Services", which was based on this idea.
2. In some cases, doing small simple things at large scale can be profound. For example, one could take an analysis that runs on a desktop and scale it out to run 100 to 100,000 times at once. In this way, the small science that is done in the long tail becomes large science.
3. Layers can provide a scaling mechanism, even if they are not simple. Different layers can aggregate and funnel needs from layers above to layers below. In this model, as one goes up the stack, things become more customized.

### 3c - Grand Challenge Communities

**How to develop the "blueprint"?**

One of the most appropriate means of developing a blueprint is to have the global scientific community agree on interfaces, and then to have the community compete on implementation. Data and software, as well as research results, seem to be drivers of developing the blueprint. But it is important to link those three with each other. Publications of research results ought to be linked together with data sources and with software. And all of these should be coordinated internationally, either by joint meetings of grand challenge science teams that would decide on global data policies and their implementations, and/or by Global Software Institutes and Centers of Excellence for each scientific discipline. There is a need for "shepherds of global codes" who would provide particular software services to global scientific community.

# 4 - Immediate actions

The three breakout groups were challenged to think of a few things that could be done relatively quickly to make a difference.

## 4a - Big Science

There were no specific recommendations from the Big Science breakout – apart from support the need to establish a blueprint process.

## 4b - Small Science / Long Tail

1. Can be done quickly: Having seen the presentation from Ed Seidel on Big Science, including how it is changing and what is going to be needed to enable continuing scientific progress, we think **a similar presentation is needed that focuses on long-tail science**. The purpose of this is to inspire the audience, and to create a shared vision of what is needed that the audience will want to work towards.

2. Can be started quickly, completed in six months: There is already a lot of good work being done by scientists in the long tail. **We need to survey what they are doing**, focusing on the solutions that work, analyzing why they work, and understanding what is common among groups. This could also be used to find elements (software, infrastructures, tools, services) that should be provided more broadly.

3. Medium term: **We want to provide a simple-to-use workflow environment that supports easy definition of simple workflows**. This needs to be able to move data, manipulate data, analyze/process data, etc. It needs to work with applications and tools that the scientists already use. It might be an abstraction of tools that already exist for some long-tail researchers, such as science gateways, Galaxy, Robetta, etc.

4. Longer term: **A pervasive storage service is needed**, so that scientists can store data, knowing that it will be safe, and knowing that it will be easy to access from a variety of places, including desktops/labs and complex components of cyberinfrastructure.

### 4c - Grand Challenge Communities

Group members believed that an international charrette-like process should be launched by a fairly small, but representative steering committee to define a blueprint for global CI. Consulting with all interested stakeholders is very important for the success of the process. EarthCube was mentioned as an example of a successful charrette process, although its scope was narrow – the US region only. The goal of EarthCube is to transform the conduct of research by supporting the development of community-guided cyberinfrastructure to integrate data and information for knowledge management across the Geosciences. We need something similar for the global CI and all grand challenge community science fields.

# 5 - Workshop outputs: mission statement and blueprint process

A mission statement (§5a) was collaboratively written at the end of the workshop, as a statement of what the attendees felt was agreed upon during the meeting. In addition, the authors have tried to summarize the overall consensus of the meeting in a proposal for a blueprint process (§5b), which also captures the discussion in a plenary session during the workshop.

## 5a - Mission Statement, version 1.0

E-Science is increasingly based on global collaborations that need access to different resources (e.g., people, instruments, libraries, data, compute, software, algorithms, etc.) across different science communities (e.g., Research Infrastructures, Grand Science Challenges, Societal Challenges), which are at different levels of maturity and, of different sizes and impact (e.g., niche high-end science, campus science, citizen science, etc.). To deliver sustainable 'infrastructure' across all of these areas it is necessary to engage with representative stakeholders from across all of these areas to understand their usage scenarios and resulting requirements from which a multi-year plan that identifies and delivers the operation, maintenance and development of the key common components that can be agreed across the communities.

To sustainably exploit the successful prototyping and experimental work that has taken place over the last decade, it is necessary to understand how to effectively identify and deliver the services that need to continue to be operated and identify the new services that need to be provisioned to support the increasingly data driven science.

Therefore, the activity coming out of this workshop will focus on establishing an on-going integrated multi-disciplinary and global view on future infrastructure requirements by:

- Establishing a globally coordinated process/framework for the development, integration, operation and maintenance of these common (across region and discipline) infrastructure components.
- Establishing a responsive requirements-driven process that identifies the common infrastructure components that are needed.

The ongoing process requirement is to:

- Collect usage scenarios
- Distill requirements
- Identify commonalities
- Assemble common components into an 'architecture' that can be established

A sustainable future for the digital science community that allows for maximal science return through easy controlled sharing between domains requires integration globally between:

- E-Infrastructure & cyberinfrastructure providers
- The dynamic provisioning of high-performance networking
- The services and tools needed to provision and use the infrastructure
- The data services of curation, discovery and movement
- The applications and their algorithms needed to support the science

Each of these areas will continue to evolve along different timelines and costs associated with them and to ensure long-term sustainability for what will be a multi-year (decade?) plan where continuity and vision is essential, engagement with policy makers and funders.

Moving forward we have to figure out how to:

1. Maintain and evolve the infrastructure
2. Develop more cost effective ways to do it
3. Identify needs and commonalities

## 5b - Proposal for a blueprint process

The mission as stated above is expected to be implemented in continuous process (the 'blueprint process'). That process should not be bound to a specific (set of) CI instances, and should be open to all CI stakeholders (providers, operators, consumers, funders). Science community involvement is essential to the process, but also difficult (hard to balance communities, what is the value proposition for participating).

The motivation for this process is:

- to provide sustainability to CI, and to ensure funding is well-spent and well-directed
- to support new international grand challenge communities as well as ongoing science projects
- to provide effective international cooperation
- to provide a foundation for news aspects in data and software

The process will include elements such as:

- security assurances, identity management
- data sharing policies
- collaborative software support
- campus bridging to international infrastructures
- governance mechanisms
- best practices
- reuse
- continued innovation

It will be a continuous challenge to balance the stability of the blueprint, as a well defined layered stack of interfaces, with the necessity to evolve CI along with changing science requirements and changing technologies. Thus there will not be one blueprint, but rather a blueprint **process**.

We propose that the process should be driven (not dominated) by a small, focused steering group, and should follow a regular meeting cadence. Support from the funding agencies will be essential, not only from the practical point of view (travel funding etc.), but also from the process point of view (charrettes). Also, the agencies will need to implement parts of the blueprint recommendations in order make it useful; failing that, the blueprint will remain an academic exercise. Finally, the funding agencies will have to support the process by encouraging and supporting the participation of science communities.

The blueprint will represent an architecture template for CI, likely rendered as a layered stack of interface definitions. As the blueprint will need to continuously evolve, the output of the process will be snapshots of that blueprint.

As mentioned, the blueprint will likely be rendered as a layered stack of interfaces (vs. a stack of implementations and technologies); it thus seems prudent to assume that the blueprint can benefit from, but also drive standardization efforts.

It is an option to use the infrastructure and process of the OGF [8] to implement the blueprint process (mailing list, wiki, meeting cadence, etc.). Establishing an independent process is obviously also possible.

# 6 - Summary

The workshop "Grid Computing: The Next Decade" is documented through this report. We believe that the workshop and this report make a significant contribution toward establishing a common vision among the community leaders who have to do the work to unify the field, and among the government agencies that are asked to fund the activities. There is still a long way to go, especially when it comes to international cooperation between funding agencies and scientific infrastructure providers, but the organizers and participants of the workshop believe that this is an important goal that should be achieved. Many fields have depended on computational science simulations, and many now are beginning to depend on computationally intensive data analysis. Infrastructure providers seek to build computational systems that support these researchers. Developing the common vision that is needed to support these efforts is the eventual goal of this activity. The meeting brought international experts representing such distributed scientific infrastructures as XSEDE, EGI, OSG, NAREGI, PRACE, FutureGrid, and PL-Grid, among others, but even broader participation in future activities is needed.

# 7 - Next Steps

The workshop organizers have compiled this report with the hope of receiving further comments from (a) other workshop participants, and (b) the broader eScience community. We are aware that the workshop attendees were mostly infrastructure providers, and while many participants have applications experience, only a few could claim to represent an application community.

The workshop was a follow-up to the SC 2011 BOF "Towards a Unified Cyberinfrastructure" [9] and it is part of a series [10]; we strongly encourage the community to watch our activities and participate in the future workshops and ongoing discussion. At this point we are seeking for comments on this document, especially the mission statement and the process presented above. All CI stakeholders are invited! Please send your comments to: grid-next-decade@lists.man.poznan.pl

# 8 - References